\newcommand{\teff}{T$_{\mathrm{eff}}$}

\newcommand{\logg}{log $g$}
\newcommand{\feh}{[Fe/H]}
\newcommand{\kms}{km~s$^{-1}$}

\newcommand{\vmicro}{$\xi$}

\newcommand{\project}[1]{\textsl{#1}}
\newcommand{\gaia}{\project{Gaia}}

\documentclass[a4paper,fuseAMS,leqn,usenatbib]{mnras}

\usepackage{graphicx}
\usepackage{amssymb}
\usepackage{amsmath}
\usepackage{times}
\usepackage{subfigure}
\usepackage[T1]{fontenc}
\usepackage{ae,aecompl}

\title[Chemically Tagging HVS]{The Fastest Travel Together: Chemical Tagging of the Fastest Stars in Gaia DR2 to the Stellar Halo}

 \author[Hawkins \& Wyse 2018]{Keith~Hawkins$^{1}$\thanks{E-mail: keithhawkins@utexas.edu} and Rosemary F. G. Wyse$^{2, 3}$ \\
$^{1}$Department of Astronomy, The University of Texas at Austin, 2515 Speedway Boulevard, Austin, TX 78712, USA \\
$^{2}$Physics and Astronomy Department, Johns Hopkins University, 3400 North Charles Street, Baltimore, MD 21218, USA\\
$^{3}$Leverhulme Trust Visiting Professor, University of Edinburgh, UK
 }

\date{Accepted 2018 August 13. Received 2018 August 12; in original form 2018 June 14}

\pubyear{2018}

\begin{document}
\label{firstpage}
\pagerange{\pageref{firstpage}--\pageref{lastpage}}
\maketitle

\begin{abstract}
The fastest moving stars provide insight into several fundamental properties of the Galaxy, including the escape velocity as a function of Galactocentric radius, the total mass, and the nature and frequency of stellar encounters with the central Supermassive Black Hole. 
The recent second data release of \gaia\  has allowed the identification of new samples of stars with extreme velocities. Discrimination among the possible origins of these stars is facilitated by chemical abundance information. We here report the results from our high-resolution spectroscopic followup, using the Apache Point Observatory, of five late-type `hypervelocity' star candidates from \cite{Marchetti2018}, characterised by derived total Galactic rest-frame velocities between 500--600~\kms\ and estimated, by those authors, to have a  probability larger than 50\% to be unbound from the Milky Way. Our new results confirm the \gaia\ DR2 radial velocities to within 1~\kms. We derived stellar atmospheric parameters and chemical abundances for several species including  $\alpha$-elements (Mg, Ti, Si, Ca), Fe-peak elements (Fe, Ni, Co, Cr, Mn), neutron-capture elements (Sr, Y, Zr, Ba, La, Nd, Eu) and odd-Z elements (Na, Al, K, V, Cu, Sc).  We find that all stars observed  are metal-poor giants with --2 $\leq$ \feh\ $\leq$ --1~dex and are chemically indistinguishable from  typical halo stars. Our results are supported by the chemical properties of 
four additional stars with extreme space motions which were observed by existing spectroscopic surveys. We conclude that these stars are simply the high-velocity tail of the stellar halo and effectively rule out more exotic origins such as from the Galactic centre or the Large Magellanic Cloud.
\end{abstract}


\section{Introduction}
\label{sec:Introduction}
The fastest moving stars observed within the Galaxy provide insight into several fundamental properties of the Galaxy, including the escape velocity (potential well depth) as a function of Galactocentric radius and the total mass \citep[e.g.][]{Piffl2014, Rossi2017}, and the nature and frequency of stellar encounters with the central Supermassive Black Hole \citep[e.g.][]{Rossi2017}. Despite their importance, the nature of these rare objects and their ejection mechanisms remain poorly understood \citep[reviewed in][]{Brown2015}.

The existence of  `hypervelocity' stars (HVSs)\footnote{In this study we define hypervelocity stars as those whose velocity is estimated to be larger than the escape speed and thus unbound. This definition is agnostic to the production mechanism.}, moving at speeds in excess of 1000~\kms, was first predicted by \cite{Hills1988}, being produced as a result  of triple-body encounters between binary stellar systems and the putative supermassive black hole (SMBH) at the Galactic Centre.  They remained a theoretical fascination until the early 2000s when \citet{Brown2005} serendipitously discovered  a 3~M$_{\odot}$ B-type star at a distance of 100~kpc and with a Galactic rest frame radial velocity above 700~\kms, indicating that it was unbound to the Milky Way.  The advent of several large spectroscopic and astrometric surveys has greatly increased the potential of finding and characterising these interesting rare stars has led to significant growth in this subfield. 

Over the last decade, there has been significant interest in understanding whether other production mechanism could be responsible for the origin of HVSs. Several mechanisms other than the `Hills' mechanism have been proposed. These include the ejection of a binary companion when the primary star undergoes a supernova \citep[e.g.][]{Blaauw1961}, tidal debris from dwarf galaxies \citep{Abadi2009}, dynamical ejection from dense stellar clusters \citep{Poveda1967, Bromley2009}, acceleration from  the jets of Active Galactic Nuclei \citep[e.g.][]{Silk2012,Wang2017}, among others. The various different production mechanisms will have different observational signatures in the spatial, kinematic, and chemical distributions. However, there are only $\sim$20 confirmed HVSs.

With the recent second data release of data from the \gaia\ astrometric mission (\gaia\ DR2), there has been a concerted effort to search for the fastest moving stars in the Galaxy, motivated by a range of scientific questions. For example, \cite{Shen2018} report the discovery of three white dwarfs with derived space motions over 1000~\kms, plausibly survivors of double-detonation events.  \cite{Boubert2018} examine the status of hypervelocity stars, with an emphasis on late-type stars. Even though many candidates analysed by \cite{Boubert2018} are not likely to be actually unbound stars, the new data have enabled the identification of many new candidate extreme-velocity stars at a higher level of significance.  \cite{Marchetti2018} and \cite{Hattori2018} both undertake a systematic search for the fastest moving stars in \gaia\ DR2, using the total velocity vector of a star, or just its Galactic rest frame tangential motion, respectively. The Gaia data also allow the orbits of extreme-velocity stars to be derived, once a model for the Galactic potential is adopted  \citep[e.g.][]{Brown2018}. These recent studies make it clear that there has been a recent rapid increase in the number of candidate late-type extreme velocity stars identified and characterising their orbital {\it and} chemical properties will be critical to understand their origins.

The discovery of new late-type candidate extreme-velocity stars from \gaia\ DR2 warrants spectroscopic followup. The purpose of this work is to communicate the results from our high-resolution spectroscopic followup of five late-type candidates from the sample of extreme-velocity stars identified by  \cite{Marchetti2018}. The aim of this paper is two fold: (i) confirm (or not) the radial velocity (RV) reported by \gaia\  \citep[many have very high RVs, which  the \gaia\ team caution against using][]{Gaiasummary2018}, and (2) provide detailed characterization of these stars in terms of both stellar atmospheric parameters and chemistry,  in order to better understand their origins. 

With these aims in mind, this paper is arranged as follows: in
section~\ref{subsec:selection}, we review the selection criteria of
HVS candidates using \gaia\ DR2, which were defined in
\cite{Marchetti2018}. We then outline the high-resolution
spectroscopic followup that we completed
(section~\ref{subsec:APOfollow}) and our search to find if any of the
HVS candidates have serendipitously been observed by any of several large
spectroscopic surveys with publicly available data
(section~\ref{subsec:largespec}). In section~\ref{sec:method} we
describe the methods that were used to derive the RV, stellar
parameters and chemical abundances for 22 species. The results
confirming the RVs and determining the stellar parameters and chemical
abundances in 4 elemental families ($\alpha$, odd-Z, Fe-peak, neutron
capture) are described in section~\ref{sec:results}. These results are
then placed into the context of how HVSs may have formed in
section~\ref{subsec:disucssion}. Finally, we summarize our results in
section~\ref{sec:summary}.

\section{Data} \label{sec:data}
\subsection{Selecting Extreme Velocity  Stars from Gaia DR2}
\label{subsec:selection}
There are various possible criteria and selection techniques that may be used to identify candidate extreme-velocity stars. Prior to the precise astrometric data from \gaia, a high value of the Galactic rest-frame radial velocity was used \citep[e.g.][]{Brown2005, Smith2007, Piffl2014, Hawkins2015}. The use of high proper-motion to identify `high-velocity' (relative to the Sun) halo stars has a long history \citep[e.g.][]{Eggen1962, Carney1994}. The addition of parallax information allows the adoption of high tangential velocity in the Galactic rest frame as the criterion, as utilised by \cite{Hattori2018}. The addition of radial velocities allows the adoption of high 3-dimensional space motion, as utilised by \cite{Marchetti2018}. We here select stars from the extreme-velocity sample defined by \cite{Marchetti2018} and we review their procedure next.

\cite{Marchetti2018} first divided the over-seven million stars in the \gaia\ DR2 catalogue with both radial velocity and 5-parameter astrometric solution into a high-quality subsample (with fractional parallax uncertainties below 20\%, (6376803 stars) and a low-quality subsample (806459 stars). They derived distances for the high-quality subsample by simple inversion of the parallax, while for  the low-quality subsample distances were inferred from the parallax using the procedure of \cite{Astraatmadja2016}. The 3-d space-velocity vectors for stars in both subsamples were derived using  the full covariance matrix in proper-motion space. These authors then implemented some quality-control cuts (see their section~4), to ensure, for example, that the astrometric model was a good fit to the data obtained from \gaia. They then applied  a minimum velocity threshold such that the absolute value of the 3-d Galactic rest-frame velocity be larger than 450~\kms. This resulted in a sample of 165 high-velocity stars, most of which lie on the red-giant locus in the HR~Diagram and are located a few~kpc from the Sun. \cite{Marchetti2018} then adopted a model Galactic potential (see their section 4.1 for details) and identified 28 extreme-velocity stars that they estimated to have a greater than 50\% probability of being on unbound orbits. These authors further used the derived orbit to infer whether a given star could have originated in the Galactic Centre or from a location within the disc. We have drawn our sample for high-resolution spectroscopic followup from this final set of stars (their Table~1).

\begin{table*}
\caption{Observational Properties of Candidate Extreme-velocity Stars and Radial Velocity  Standards}
\label{tab:obsprops}
\centering 
\begin{tabular}{l l l l l l l l l l l   }
\hline\hline
Star & RA & DEC & $G$ & RV$_{Gaia}$ & RV$_{\mathrm{APO}}$ & V$_{tot}$ & P$_{MW}$ & P$_{ub}$ & SNR & t$_{obs}$ \\
 & (deg) & (deg) & (mag) & (\kms) & (\kms) & (\kms) &  &  & & (s) \\
 \hline
 1508756353921427328 & 210.2049 & 45.86615 & 12.63 & 128.22$\pm$1.3&122.68$\pm$0.30 & 515 & 0.99&0.58& 44 & 1800\\
 1364548016594914560 & 268.7792 & 50.57305 & 11.93 & 110.36$\pm$44 & 109.50$\pm$0.20 & 531 &1.00 & 0.74 & 46 & 1800 \\
 4593398670455374592 & 274.8965 & 33.81894 &12.24 & -312.96$\pm$1.15 & -314.60$\pm$0.23& 545 & 0.55 &0.58&42 & 1800\\
 4248140165233284352 & 299.6680 & 04.51105 & 13.21 & -358.10$\pm$2.28&-358.29$\pm$0.17&567 & 0.50&0.56&42 & 1800 \\
 2233912206910720000 & 299.2838 & 55.49696 & 12.97 & -343.93$\pm$1.71&-344.19$\pm$0.50& 540&0.42&0.76$^a$ & 34 & 1800\\

 \hline

\hline
 &  &  &   & RV standards & &  &  & & \\
 \hline
 HD~122563&210.6318 & 9.68579 & 5.85 & -26.16$\pm$0.16 & -26.57$\pm$0.64 & ... & ... & ... & 210 & 60 \\
 HD~84937 &147.2354 & 13.7409 & 8.19 & -15.69$\pm$0.42 & -15.50$\pm$0.50 & ... & ... & ... & 150 & 600\\

\hline  \hline     
\end{tabular}
\\
\raggedright
NOTE: The identifier of each star is given in
column~1, with coordinates in columns~2 and 3.  The $G$-band magnitude
is tabulated in column~4. The radial velocity measured by \gaia\ and by our high-resolution spectra are in
column~5 and column~6, respectively, with the 3-d speed in column~7.   The probabilities of the star originating  in the Milky
Way disc (P$_{MW}$) and of being unbound to the Galaxy (P$_{ub}$) are
taken from \cite{Marchetti2018} and given in columns~8 and 9 respectively. $^a$ \cite{Hattori2018} derive a probability of 0.84 for this star being unbound from the potential they adopt.  Columns~10 and 11 give, respectively, the signal-to-noise ratio measured in the $\lambda \sim$~5300--5400~\AA\ range and the exposure time in seconds required to obtain the  signal-to-noise ratio.

\end{table*}

However, it is critical to keep in mind that there are other valid procedures for selecting HVS candidates. For example, \cite{Hattori2018} search for HVS candidates not from full 3-d space motions as in \cite{Marchetti2018} but rather  using the Galactic rest frame tangential velocity corrected for the Solar reflex motion. They found a total of 30 extreme-velocity stars. Another way is to use the Galactic rest-frame radial velocity to select high-velocity stars \citep[e.g.][]{Hawkins2015}. Each of these methods is valid and should be combined to find the best candidates for unbound HVSs. 

\subsection{Observational Follow-up and Reduction} \label{subsec:APOfollow}

\begin{figure*}
	 \includegraphics[width=2\columnwidth]{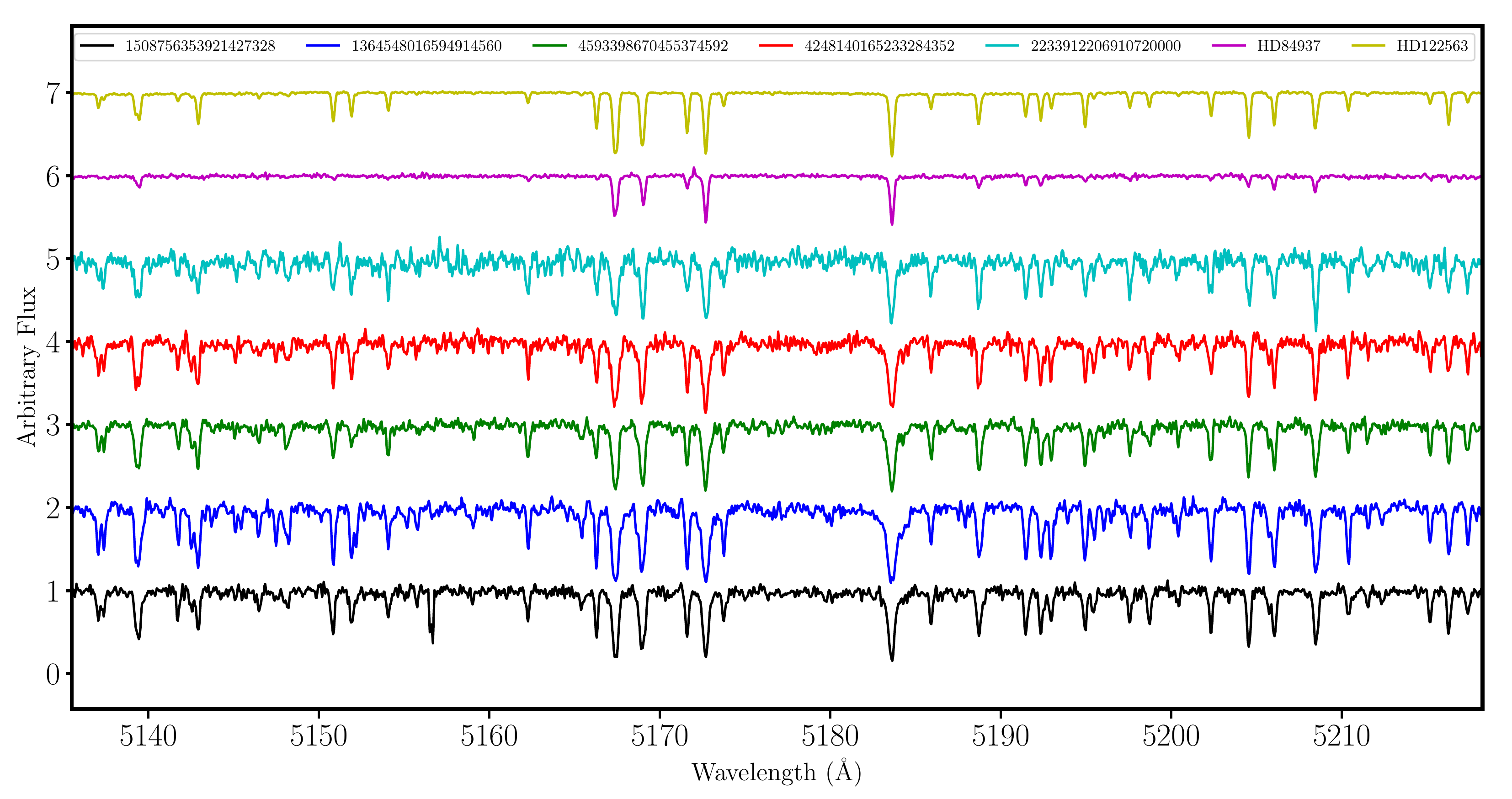}
	\caption{Illustrates the observed spectra in the Mg triplet region (5135--5215~\AA) of the two bright standard stars (HD~122563, HD~84937) and the five  candidate HVSs observed in this work, namely  \gaia~DR2 1508756353921427328, \gaia~DR2 1364548016594914560, \gaia~DR2 4593398670455374592, \gaia\ DR2 4248140165233284352 and \gaia~DR2 2233912206910720000  from top to bottom, respectively.} 
	\label{fig:spectra}
\end{figure*}

We selected five candidate extreme-velocity stars, accessible  from Apache Point Observatory (APO) in May~2018, from the sample of \cite{Marchetti2018}. Of these five, three have derived orbits consistent with an origin in the Galactic centre while the other two do not. We obtained high-resolution (R $=\lambda/\Delta\lambda \sim$ 31500) optical spectra for each of these stars, using the ARC Echelle Spectrograph (ARCES) on the APO 3.5m  telescope, during  the nights of 2-May-2018 and 7-May-2018. The spectra have wavelength coverage  3800--9200~\AA\ over $\sim$107 echelle orders.  We also obtained spectra for two bright radial-velocity standard stars (HD~84937 and HD~122563). Standard calibration images (biases, flat fields, and ThAr arc lamps) were taken. The spectra were reduced using the standard methods of: bias subtraction, extraction, flat field division, scattered light subtraction, and stacking using the echelle package of IRAF\footnote{Distributed by NOAO, operated by AURA under cooperative agreement with the NSF.}. An initial continuum normalization for order stitching was done assuming a 5th-order spline function. RVs were measured using cross-correlation with an Arcturus spectral template. When multiple spectra of the same object were taken, the individual spectra were co-added to obtain a higher signal-to-noise ratio (SNR).

The final high-resolution spectra of the five extreme-velocity targets have a typical SNR $\sim$ 40 pixel$^{-1}$.  The basic observational properties of these stars (plus the standard stars) are given in Table~\ref{tab:obsprops}, including coordinates, \gaia\ G band magnitude, radial velocity as measured in both \gaia\ and this study, and the typical SNR.  Figure~\ref{fig:spectra} shows the final reduced, extracted, wavelength calibrated, and RV corrected spectra in the Mg-triplet region (5135--5215~\AA). These spectra were used to carry out our scientific aims, namely to confirm (or not) the \gaia\ radial
velocities, characterise the atmospheric parameters and in particular the chemical properties of these rare stars and hence investigate how they might be associated, via `chemical tagging',  with known populations within the Milky Way or LMC.

\subsection{Large Spectroscopic Surveys : Data From APOGEE, LAMOST, and RAVE} \label{subsec:largespec}
We cross matched the extreme-velocity candidate sample from Table~1 of \cite{Marchetti2018} with the public databases of four large spectroscopic surveys, namely APOGEE DR14 \citep[Holtzman et al., in prep.]{Majewski2017}, LAMOST DR3 \citep[]{Luo2015, Xiang2017}, RAVE DR5 \citep{Kunder2017}  and GALAH DR2 \citep{Buder2018}. We found 1 star (Gaia DR2 1268023196461923712) in common with APOGEE, 3 stars (\gaia~DR2 6639557580310606976, \gaia~DR2 5212817273334550016, and  \gaia~DR2 4916199478888664320) in common with
the RAVE survey, 1 star (Gaia~DR2 1268023196461923712) in common with the LAMOST survey and no stars in common with the GALAH survey. The basic observational data for these stars are given in Table~\ref{tab:obspropsothers}. 

%
%
%
%

\begin{table*}
\caption{Observational Properties of Extreme-velocity Star Candidates from other surveys}
\label{tab:obspropsothers}
\centering 
\begin{tabular}{l l l l l l l l l l l  }
\hline\hline
Star & RA & DEC & $G$ & RV$_{Gaia}$ & RV$_{\mathrm{survey}}$ & V$_{tot}$ & P$_{MW}$ & P$_{ub}$ & SNR & Survey \\
 & (deg) & (deg) & (mag) & (\kms) & (\kms) & (\kms) &  &  & \\

\hline 
\hline

 1268023196461923712 & 225.7835 & 26.24632 & 13.00 &  -276.84$\pm$1.64 & -275.49$\pm$0.03$^a$& 551&0.49&0.62&119&A, L\\ 
 6639557580310606976 & 287.9278& -57.80599 &  12.26& -257.08$\pm$1.19 & -255.59$\pm$0.75&581 &0.38&0.62&63&R\\
 5212817273334550016 & 107.1991 & -76.21933 & 10.89 & 159.87$\pm$0.29 & 157.97$\pm$0.44& 568&0.35&0.79&81&R\\
 4916199478888664320 & 23.38253 & -51.92318 & 12.61 & 86.87$\pm$1.33 & 86.36$\pm$2.15& 538&0.44&0.67&...&R\\

\hline  \hline     
\end{tabular}
\\
\raggedright
Columns are as Table~\ref{tab:obsprops}, with the addition of the Survey identifier in column~11. These are designated with A for APOGEE DR14 \citep[][Holtzman et al., in prep.]{Majewski2017}, L for LAMOST \citep[][]{Luo2015, Xiang2017}, and R for RAVE \citep{Steinmetz2006, Kunder2017}.  $^a$The RV for Gaia~DR2~1268023196461923712 comes from the APOGEE survey (that from the LAMOST survey is  consistent to within a few \kms).
\end{table*}

\section{High-resolution Spectroscopic Analysis} \label{sec:method}
Spectroscopic analysis was done using the Brussels Automatic Code for Characterizing High accUracy Spectra \citep[BACCHUS,][]{Masseron2016}. The current released version makes use of the MARCS model atmosphere grid \citep{Gustafsson2008}, and the radiative transfer code TURBOSPECTRUM \citep{Alvarez1998, Plez2012} to generate synthetic spectra for comparison with the observations. Atomic lines are sourced from the fifth version of the Gaia-ESO linelist (Heiter et al., in preparation). In addition to the atomic lines, linelists for molecular species are also used in the calculation of synthetic spectra. The molecular species included are CH \citep{Masseron2014}, and CN, NH, OH, MgH and  C$_{2}$(T. Masseron, private communication); the lines of SiH molecules are adopted from the Kurucz linelists and those from TiO, ZrO, FeH, CaH from B. Plez (private communication).

The BACCHUS package derives the stellar atmospheric parameters - effective temperature (\teff), surface gravity (\logg), microturbulent velocity (\vmicro), and iron abundance (\feh\footnote{Chemical abundances are represented in the standard way as a logarithmic ratio of element X to element Y, relative to the Sun, [X/Y], such that  $\textup{[X/Y]} = \mathrm{log}\left ( \frac{N_{X}}{N_{Y}} \right )_{star} - \mathrm{log}\left ( \frac{N_{X}}{N_{Y}} \right )_{Sun}$, where $N_{X}$ and $N_{Y}$ are the number of element X and element Y per unit volume respectively.}), under the assumption of local thermodynamic equilibrium (LTE)\footnote{Possible non-LTE (NLTE) effects on the derived values of the atmospheric parameters and chemical abundances are not taken into account.} and using the standard Fe-Ionization-Excitation equilibrium technique. Under this procedure the value of  \teff\ is derived by ensuring that there is no correlation between the abundance of Fe, denoted as $\log_{10}(\mathrm{A_{Fe}}$), and the excitation potential (in eV) of the lines being used. The value of \logg\ is determined by ensuring that there is no significant offset between the abundance of neutral Fe (\ion{Fe}{I}) and that of singly ionized Fe (\ion{Fe}{II}). Finally, the value of the microturbulent velocity, \vmicro, is derived by forcing there to be no correlation between the abundance of Fe and the reduced equivalent width (REW, defined as equivalent width divided by the wavelength of the line). These steps in the derivation of  the stellar atmospheric parameters were completed using up to 90~\ion{Fe}{I} lines and 30~\ion{Fe}{II} lines. For more details about BACCHUS, we refer the reader to Section~2.2 of \cite{Hawkins2015}. 

The chemical abundances, reported as  [X/H], were derived for each element and each absorption feature by first fixing the stellar parameters to those derived as described above and synthesizing spectra with different values of [X/H]. A $\chi^2$ minimization was done between the observed spectrum and the synthesized spectra with different [X/H] values to obtain the elemental abundance. This process was repeated for each element and resulted in abundances for 22 species. We take the median and dispersion of the abundances derived from individual lines, divided by the  square root of the number of lines used, as the quoted abundance and internal error, respectively. The abundances are scaled relative to the Sun by adopting the Solar abundances for each element from \cite{Asplund2005}.

\section{Results and Discussion} \label{sec:results}

\begin{table*}
\caption{Stellar parameters and chemical abundances for 5 Candidate Extreme-velocity Stars}
\label{tab:results}
\centering 
\begin{tabular}{l l l l l l }
\hline\hline
Star & 1508756353921427328 &  1364548016594914560 & 4593398670455374592 & 4248140165233284352 & 2233912206910720000 \\
\hline
\teff\ (K) & 4699 $\pm$ 85 & 4640 $\pm$ 98 &4868 $\pm$ 65 &4718 $\pm$ 49 &5070 $\pm$ 7 \\ 
T$_{\mathrm{eff,phot}}^a$ (K)& 4967$^{+75}_{-57}$& 4813$^{+220}_{-261}$ & 5469$^{+774}_{-441}$& 4858$^{+89}_{-72}$&5158$^{+802}_{-79}$\\
\logg\ (cgs) & 0.90 $\pm$ 0.43 & 1.37 $\pm$ 0.49&1.47 $\pm$ 0.28&1.41 $\pm$ 0.36&1.30 $\pm$ 0.26\\ 
\feh & -1.93 $\pm$ 0.09 & -1.21 $\pm$ 0.12&-1.67 $\pm$ 0.13&-1.82 $\pm$ 0.15&-1.72 $\pm$ 0.16\\ 
\vmicro\ (\kms)& 2.04 $\pm$ 0.16 & 1.77 $\pm$ 0.10&1.58 $\pm$ 0.24&1.87 $\pm$ 0.18&2.28 $\pm$ 0.14\\ 
~[Na/Fe] & 0.21 $\pm$ 0.09 & -0.11 $\pm$ 0.05&0.53 $\pm$ 0.12&0.34 $\pm$ 0.02&... $\pm$ ...\\ 
~[Mg/Fe] & 0.64 $\pm$ 0.03 & 0.23 $\pm$ 0.07&0.58 $\pm$ 0.06&0.65 $\pm$ 0.13&0.31 $\pm$ 0.13\\ 
~[Al/Fe] & ... $\pm$ ... & 0.33 $\pm$ 0.10&0.07 $\pm$ 0.10&... $\pm$ ...&...$\pm$ ...\\ 
~[Si/Fe] & 0.58 $\pm$ 0.05 & 0.08 $\pm$ 0.05&0.37 $\pm$ 0.05&0.56 $\pm$ 0.05&0.77 $\pm$ 0.08\\ 
~[K/Fe] & 1.19 $\pm$ 0.10 & 0.75 $\pm$ 0.10&... $\pm$ ...&1.38 $\pm$ 0.10&... $\pm$ ...\\ 
~[Ca/Fe] & 0.50 $\pm$ 0.03 & 0.26 $\pm$ 0.04&0.50 $\pm$ 0.03&0.53 $\pm$ 0.03&0.54 $\pm$ 0.04\\ 
~[Sc/Fe] & 0.17 $\pm$ 0.05 & 0.20 $\pm$ 0.02&0.26 $\pm$ 0.04&0.24 $\pm$ 0.08&0.08 $\pm$ 0.13\\ 
~[Ti/Fe] & 0.29 $\pm$ 0.03 & 0.32 $\pm$ 0.02&0.23 $\pm$ 0.03&0.36 $\pm$ 0.04&0.16 $\pm$ 0.06\\ 
~[V/Fe] & 0.23 $\pm$ 0.03 & 0.07 $\pm$ 0.04&-0.28 $\pm$ 0.14&0.29 $\pm$ 0.10&... $\pm$ ...\\ 
~[Cr/Fe] & -0.11 $\pm$ 0.06 & 0.10 $\pm$ 0.04&-0.30 $\pm$ 0.06&-0.03 $\pm$ 0.08&0.70 $\pm$ 0.03\\ 
~[Mn/Fe] & -0.23 $\pm$ 0.09 & -0.27 $\pm$ 0.02&-0.28 $\pm$ 0.06&-0.23 $\pm$ 0.06&-0.13 $\pm$ 0.10\\ 
~[Co/Fe] & 0.12 $\pm$ 0.06 & 0.13 $\pm$ 0.03&-0.00 $\pm$ 0.03&0.31 $\pm$ 0.06&1.30 $\pm$ 0.10\\ 
~[Ni/Fe] & 0.03 $\pm$ 0.05 & -0.04 $\pm$ 0.03&0.03 $\pm$ 0.05&0.13 $\pm$ 0.04&0.15 $\pm$ 0.07\\ 
~[Cu/Fe] & -0.33 $\pm$ 0.03 & -0.38 $\pm$ 0.15&-0.99 $\pm$ 0.17&-0.19 $\pm$ 0.13&0.31 $\pm$ 0.10\\ 
~[Zn/Fe] & -0.11 $\pm$ 0.01 & -0.13 $\pm$ 0.11&0.06 $\pm$ 0.04&0.00 $\pm$ 0.01&0.16 $\pm$ 0.10\\ 
~[Sr/Fe] & 0.10 $\pm$ 0.10 & -0.41 $\pm$ 0.10&... $\pm$ ...&0.22 $\pm$ 0.10&... $\pm$ ...\\ 
~[Y/Fe] & -0.11 $\pm$ 0.08 & -0.20 $\pm$ 0.07&-0.11 $\pm$ 0.05&-0.18 $\pm$ 0.10&-0.14 $\pm$ 0.01\\ 
~[Zr/Fe] & 0.23 $\pm$ 0.13 & 0.37 $\pm$ 0.10&0.59 $\pm$ 0.13&0.90 $\pm$ 0.03&0.77 $\pm$ 0.10\\ 
~[Ba/Fe] & -0.02 $\pm$ 0.12 & 0.18 $\pm$ 0.12&0.56 $\pm$ 0.03&0.02 $\pm$ 0.13&-0.15 $\pm$ 0.15\\ 
~[La/Fe] & 0.28 $\pm$ 0.03 & 0.26 $\pm$ 0.04&0.16 $\pm$ 0.06&0.66 $\pm$ 0.07&0.71 $\pm$ 0.08\\ 
~[Nd/Fe] & 0.31 $\pm$ 0.02 & 0.28 $\pm$ 0.03&0.17 $\pm$ 0.03&0.74 $\pm$ 0.06&0.43 $\pm$ 0.08\\ 
~[Eu/Fe] & 0.77 $\pm$ 0.10 & ... $\pm$ ...&0.34 $\pm$ 0.10&0.70 $\pm$ 0.10&1.11 $\pm$ 0.29\\ 

\hline  \hline     
\end{tabular}
\\
\raggedright
NOTE:  The \teff, \logg, \feh, \vmicro, and [X/Fe] for 22 elements are tabulated here for each of the 5 stars in this work. We note that the uncertainty on each abundance represents only the internal error and is defined as the line-to-line dispersion of the abundance divided by the square root of the number of lines. When only 1 line can be measured for an element the assumed internal uncertainty is conservatively 0.10~dex. $^a$ We also report the photometric \teff\ based on photometry from the \gaia~DR2 (BP-RP) colour \citep[for more details consult][]{Andrae2018}.
\end{table*}
\subsection{Stellar Parameters and Radial Velocities} 
\label{subsec:SPresult}
It is immediately obvious from the entries in Tables~\ref{tab:obsprops} and \ref{tab:obspropsothers} that there is excellent agreement between the radial velocities reported by \gaia\ and those measured by us, and those reported by other surveys. The median offset is  less than 1~\kms\ and the dispersion is $\sim$2~\kms. This is reassuring, especially as the \gaia\ team cautions the use of radial velocities with high values \citep[e.g.][]{Gaiasummary2018}. That said, one star, \gaia\ DR2~1508756353921427328, has a radial velocity measurement from our APO
 spectrum that is 6~\kms\ different from that reported by \gaia. It is hard to quantify variability and binarity with just
 two epochs of data; this star warrants monitoring to determine its status.

\begin{figure}
	 \includegraphics[width=1\columnwidth]{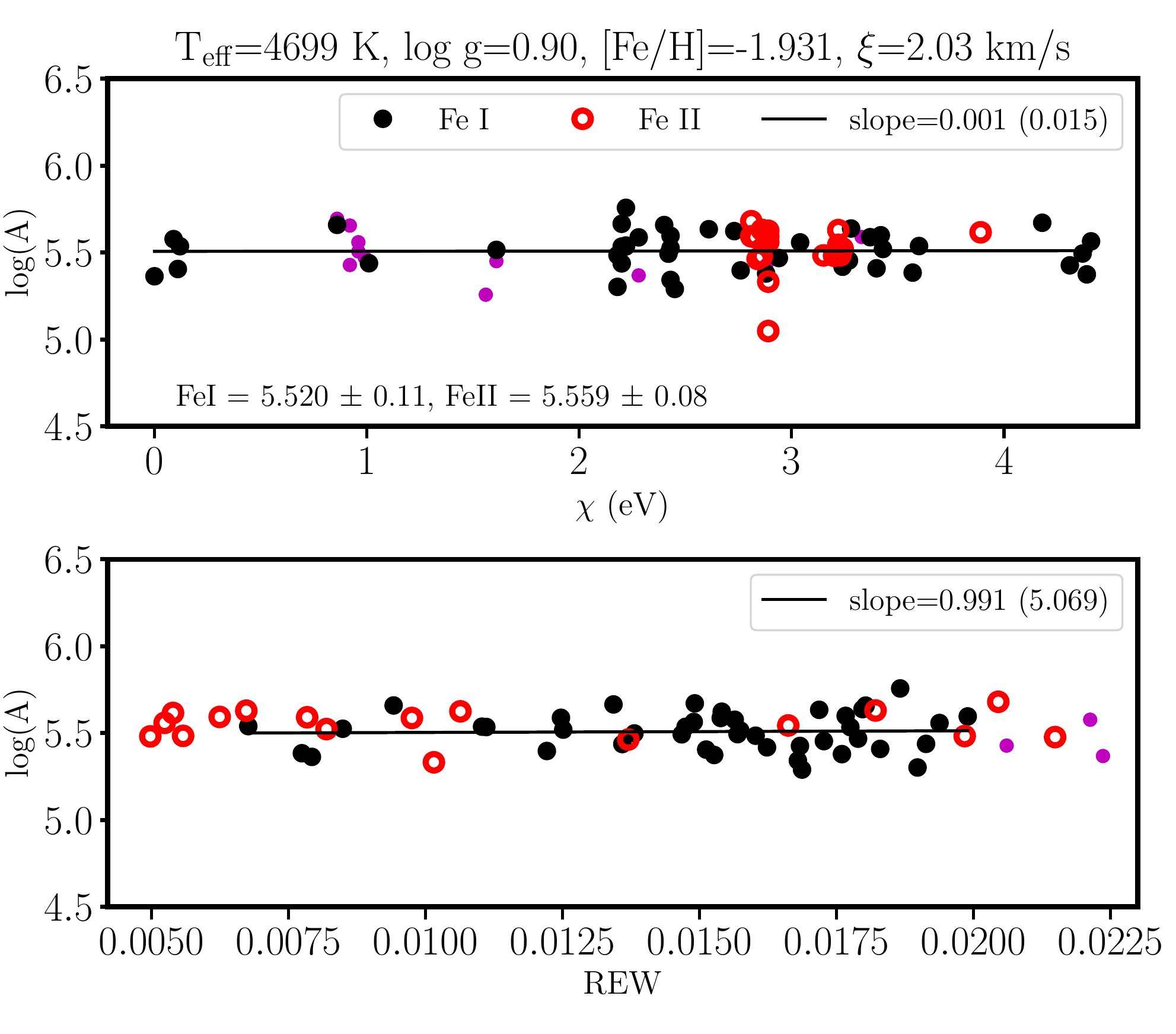}
	\caption{Illustrates the Fe excitation-ionization balance used to derive the stellar parameters for \gaia\ DR2~1508756353921427328. The upper	panel shows the log of the Fe abundance, $\log_{10}({\mathrm{A}_{\mathrm{Fe}}})$  for \ion{Fe}{I} lines (black filled circles) and
	\ion{Fe}{II} lines (open red circles) as a function of the excitation potential of the line, in eV. The lower panel shows the $\log_{10}({\mathrm{A}_{\mathrm{Fe}}})$ for \ion{Fe}{I} and \ion{Fe}{II} lines as a function of reduced equivalent width, REW = EW/$\lambda$. The magenta filled circles in both panels represent lines that are stronger than $\sim$120~m\AA\ and thus were rejected from the stellar parameter analysis. }
	\label{fig:balance}
\end{figure}

The derived values of the stellar atmospheric parameters (and the associated uncertainties) for the observed extreme-velocity candidates are given in the top 4 rows of Table~\ref{tab:results}. These parameters are derived using the standard technique of Fe ionization-excitation balance (see section~\ref{sec:method} above). An illustration of the stellar parameter solution obtained in this manner  is shown in Fig.~\ref{fig:balance} for 
star \gaia\  DR2 1508756353921427328. The upper panel of this figure shows the excitation balance where, for the adopted set of stellar parameters (given at the top of the figure), there is no correlation between the log of the \ion{Fe}{I} abundance (black filled circles) and the excitation potential. The formal slope of the linear relation fit (and its standard error) is shown in the top right corner of the panel. The black line represents the linear fit to the data.

The ionization balance, namely that the abundances derived from \ion{Fe}{I} and  \ion{Fe}{II} (red open circles) are in good agreement with each other, is also illustrated, with the actual values given in the bottom left corner of the panel. The lower panel of Fig.~\ref{fig:balance} shows the log Fe abundance for both \ion{Fe}{I} (black filled circles) and \ion{Fe}{II} (red open circles) as a function of the reduced equivalent width. This lower panel is used to constrain the value of the microturbulent velocity, \vmicro. The magenta filled circles in both panels of Fig.~\ref{fig:balance} represent very strong lines (above $\sim$120~m\AA\ EW) which are excluded from the analysis, as they may no longer be on the linear part of the curve of growth.

The resultant typical internal uncertainties of our derived  atmospheric parameters are $\sim$75~K, 0.30~dex, 0.10~dex, and 0.15~\kms\ for \teff, \logg, \feh, and \vmicro, respectively. The results from our stellar parameter analysis indicate that all of the extreme-velocity candidates observed with APO are metal-poor (--2 $<$ \feh $<$ --1.0) giant stars, consistent with their location in the colour-absolute magnitude diagram shown  in \cite{Marchetti2018}. This conclusion is also consistent with the analysis of the colour-absolute magnitude diagram of a differently defined
extreme-velocity sample by \cite{Hattori2018}. 

It is important to note that Fe abundances from metal-poor giant stars suffer from NLTE effects and these can cause the Fe excitation-ionization balance procedure to produce \teff\ values that are too small compared to other methods \citep[e.g.][and references therein]{Frebel2013}. Thus we compare our results with the photometric \teff\ values from the recent release of the \gaia\ data. These photometric \teff\ values are also tabulated in Table~\ref{tab:results}. The excitation balance \teff\ are, on average offset from the \gaia\ photometric \teff\ by --173 K (with the excitation balance \teff\ being smaller, as expected from NLTE effects) with a dispersion of 183~K. Even though the \gaia\ photometric \teff\ have very large uncertainties in some cases for these stars, we have redone the stellar parameter analysis fixing the effective temperature to the photometrically derived values from \gaia. We found that while the derived value of  \feh\ can change by as much as +0.30~dex, we are still led to the conclusion that these HVS candidate stars are metal-poor (\feh\ $<$ --1.0~dex) giants. 

As noted earlier, in addition to the extreme-velocity candidates we observed with APO, we identified four other candidates from \cite{Marchetti2018} in the public databases of  three large spectroscopic surveys (APOGEE, RAVE, and LAMOST) and thus these stars also have derived stellar parameters. Specifically, \gaia\ DR2~1268023196461923712 has additional spectra from both the APOGEE and LAMOST surveys. The published radial velocity for this star in each of the two surveys agree to within a few \kms\ and both are in good agreement with that of \gaia. The LAMOST survey reports that this star has \teff\ = 4893$\pm$10~K, \logg\ = 2.04$\pm$0.01~dex and  \feh\ = --1.58$\pm$0.01~dex while the  APOGEE survey reports  that the same star has \teff\ = 4924$\pm$100~K, \logg\ = 2.38 $\pm$0.11~dex and \feh\ = --1.62 $\pm$0.07~dex. The three
remaining stars were serendipitously observed by the RAVE survey. All three have RAVE radial velocities that agree, to within a few \kms, with the
values from \gaia. Only two of the three (Gaia DR2 6639557580310606976 and Gaia DR2 5212817273334550016) have published values of the 
stellar parameters: both are metal-poor giant stars with \feh\ = --1.49$\pm$0.17  for \gaia\ DR2 6639557580310606976  and \feh\ =  --1.80$\pm$0.14~dex for  \gaia\ DR2 5212817273334550016.

Overall, our analysis shows that these extreme-velocity candidates, selected from \cite{Marchetti2018}, are metal-poor giant
stars, with iron abundances typical of the stellar halo, albeit moving with higher than typical halo velocities. With the stellar parameters in hand, we next turn to determination of the chemical abundance patterns, treating the different elemental families in turn:  $\alpha-$elements (section~\ref{subsec:alpha}), odd-Z elements  (section~\ref{subsec:oddz}), Fe-peak elements (section~\ref{subsec:Fepeak}), and, finally,  neutron-capture elements (section~\ref{subsec:ncap}). This exploration should help discern if these stars have originated in the Galactic centre, the LMC, the Galactic disc(s) or are indeed simply part of the stellar halo.

 \subsection{$\alpha$ elements: Mg, Ti, Si, Ca} \label{subsec:alpha}

\begin{figure*}
 	\includegraphics[width=2\columnwidth]{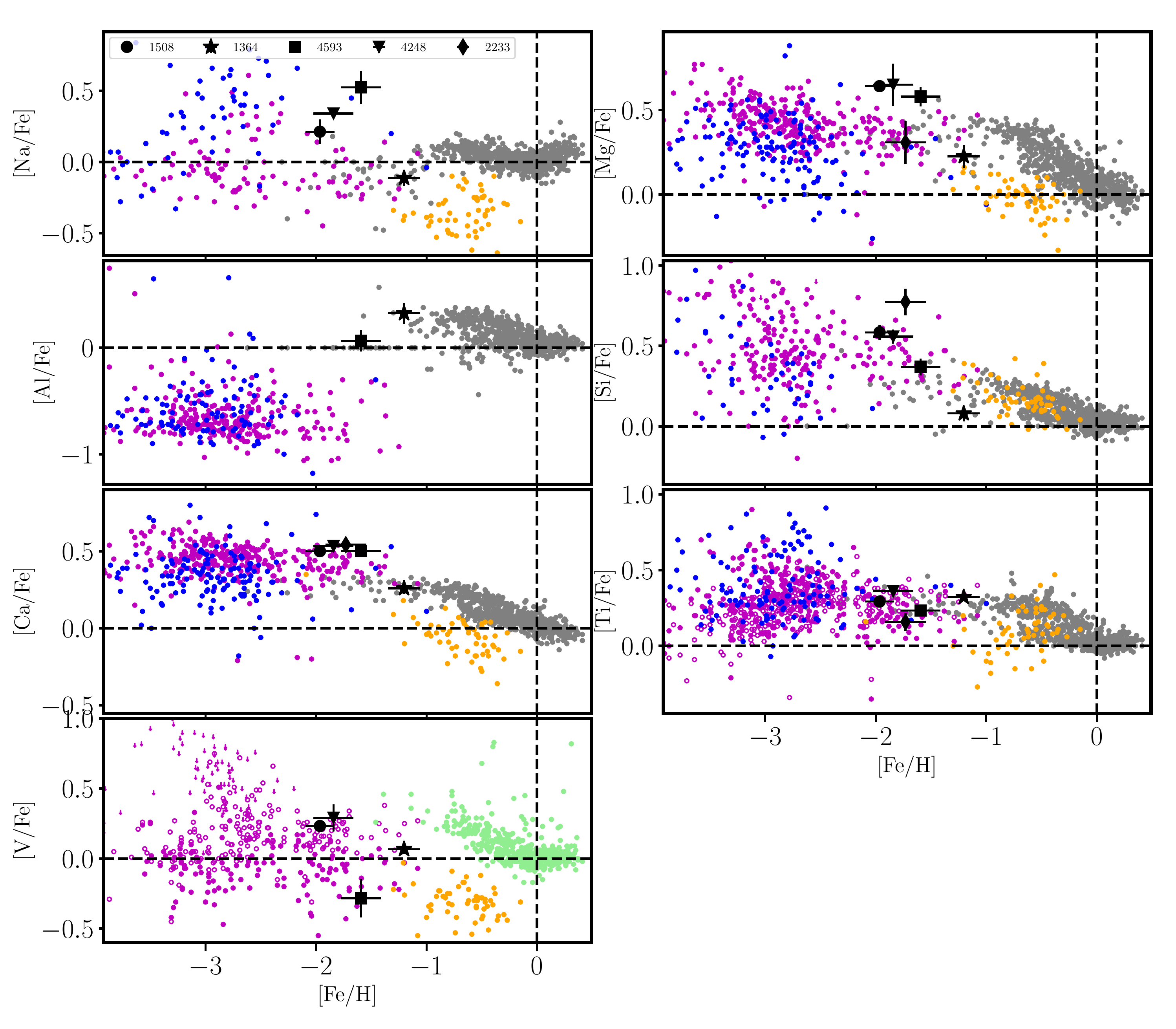}
	\caption{Illustrates the [X/Fe] ratios of the program stars (black filled symbols) as a function of \feh\ for Na, Al, Ca, and V from top to bottom, respectively, on the left and Mg, Si, and Ti from top to bottom, respectively, on the right. These elements largely represent the $\alpha$ and odd-Z elements.  The background displays abundances for the disc (including thin and thick disc) from \citet[gray filled circles]{Bensby2014} and \citet[light green]{Battistini2016}. In addition, abundances in each chemical species (where defined) for the halo are also shown, taken from \citet[magenta filled circles for neutral lines, and magenta open circles for ionized lines,]{Roederer2014} and \citet[blue filled circles]{Yong2013}. We also plot the chemical abundances from stars in the inner disc and/or bulge area of the LMC from \citet[][orange filled circles]{VanderSwaelmen2013}. Arrows represent upper limits and are shown from the results of \citet{Roederer2014}.}
	\label{fig:chem1}
\end{figure*}

The $\alpha$ elements are those which are formed via successive addition of helium nuclei ($\alpha$-particles) during the later stages of (quasistatic) nuclear fusion in the inner regions of evolved massive stars. These elements include Magnesium (Mg), Titanium (Ti), Silicon (Si), Calcium (Ca) and Oxygen (O)\footnote{We attempted to measure O for each star using the forbidden \ion{O}{I} line at 6300~\AA\ but in all cases the line could not be fit acceptably.}. The $\alpha$ elements are primarily dispersed into the interstellar medium (ISM) by core-collapse (Type~II) supernovae, which occur rapidly after the birth of the progenitor massive star. Core-collapse supernovae produce relatively little iron, so that early on, shortly after the onset of star formation, the ISM, and any newly formed stars, will be enriched to high values of $[\alpha$/Fe]. The explosive nucleosynthesis inherent in Type~Ia supernovae creates a high yield of iron, but only after a significant delay time following the birth of the progenitor (lower mass) stars. Stars that form later, after incorporation of iron from Type~Ia supernovae into the ISM, will show lower values of [$\alpha$/Fe].  The ratio of $\alpha$ elements to Fe is therefore sensitive to the star-formation history and past enrichment  of the gas from which the star formed \citep[e.g.][]{Gilmore1998, Matteucci2001,Nomoto2013}. 

This last point makes $\alpha$ elements key species that can allow us to distinguish the environment within which the rare extreme-velocity stars may have originated. Stars belonging to the local Galactic thin disc, thick disc and halo have different distributions in the $\alpha$ elements as a function of metallicity. Fig.~\ref{fig:chem1} shows the [X/Fe] abundances for the $\alpha$ elements, including Mg, Ti, Si, and  Ca, as a function of \feh\ for the extreme-velocity stars  observed in this work (black filled symbols, identified on the plots). For reference, we also show the [X/Fe] abundances for the local Galactic thin disc and thick disc taken \citet[gray filled circles]{Bensby2014} and \citet[light green filled circles, for Vanadium only]{Battistini2016} and  for the stellar halo, taken from \citet[blue filled circles]{Yong2013} and \citet[magenta filled circles for neutral lines, and magenta open circles for ionized lines]{Roederer2014}. 

We note here that the uncertainties in abundance ratios that are reported in Table~\ref{tab:results} are only internal. There are very likely systematic uncertainties for each element in addition to the internal uncertainties. The level of those systematics is difficult to assess with few stars observed with APO ARCES that are also in common with the literature sources used to represent the chemical composition of the stellar halo \citep{Yong2013, Bensby2014, Roederer2014}, disc \citep{Bensby2014, Battistini2016, Battistini2015}, and LMC \citep{VanderSwaelmen2013}. We assume the level of the systematic uncertainty in the abundances (specifically \feh) is on the order of $\sim$0.10~dex. This assumption is based on the typical offsets between the BACCHUS derived \feh\ and the adopted values for metal-poor (\feh\ $< -1.0$ dex) \gaia\ FGK benchmark stars from \cite{Jofre2014} and \cite{Hawkins2016a}. In addition, we have observed with APO ARCES a couple of metal-poor (\feh\ $\sim$ --0.70~dex) disc stars that are in common with \cite{Bensby2014} for an unrelated forthcoming study. We found an offset in \feh\ of $\sim$0.10~dex similar to the assumed value. 

As we concluded above, all of the candidate extreme-velocity stars observed are metal-poor giant stars with --2 $<$~\feh~$<$~--1~dex. This puts these stars near the peak of the metallicity distribution of the inner Galactic halo \citep[expected to be at \feh\ $\sim$ --1.5, e.g.][]{Chiba2000}. In addition, it is clear that in almost all $\alpha$ elements these metal-poor giant stars show enhanced ratios of $\alpha$/Fe relative to the Sun ([$\alpha$/Fe] $>$ 0), albeit with relatively large scatter. The level of enhancement in  [$\alpha$/Fe] we find is  typical of that found in the Galactic stellar halo. It is interesting to note that the least metal-poor star, \gaia\ DR2 1364548016594914560, is only marginally enhanced in Mg and Si compare to the other Ca and Ti but still falls in the scatter of typical stellar halo stars. The more metal-rich of these stars could also be consistent with being runaway stars  from the metal-weak thick disc. The additional extreme-velocity candidate from the APOGEE survey lacks individual elemental abundances but also shows overall $\alpha$ enhancement at a level typical of halo stars in that survey. 

The chemical distribution in the $\alpha$ elements effectively rules out that these stars have originated in the LMC, where the expected signature would be for the stars to be more metal-rich (--1.5 $<$ \feh $<$ 0.0~dex) and more depleted in the $\alpha$ elements \citep[e.g.][]{VanderSwaelmen2013}. Similarly, the chemical distribution in the $\alpha$ elements effectively rules out an origin in the Galactic centre, where it would be expected that the stars should be significantly more metal-rich (\feh > --0.8) and more enhanced in the $\alpha$ elements \citep[e.g.][]{McWilliam2016}. Instead the chemical distribution in the $\alpha$ elements of these extreme velocity stars appear most similar to those seen in typical halo stars.

\subsection{Odd-Z Elements: Na, Al, V, Cu, Sc } \label{subsec:oddz}
The odd-Z elements are produced in a variety of ways. As such, we will discuss each separately below.\\ 
\noindent {\it Na}: \\
Sodium (Na)  is thought to be produced both during carbon burning and through the NeNa cycle during H-shell burning in the post main-sequence phase of low- and intermediate-mass stars \citep[e.g.][]{Samland1998}. It is dispersed into the interstellar medium by both SNII and partially through the evolution of AGB stars \citep{Nomoto2013}. SNIa do not efficiently produce Na compared to Fe and thus [Na/Fe] is expected to decrease towards higher metallicities. Published results in the literature show that [Na/Fe] as a function of [Fe/H] for stars in the Galactic disc shows a banana shape, while in the Galactic halo Na becomes overabundant when the abundances are derived in LTE (blue points in Fig.~\ref{fig:chem1}). These values slightly decrease when NLTE (magenta points in Fig.~\ref{fig:chem1}) is considered, as by \cite{Roederer2014}. On the other hand, [Na/Fe] in the LMC has been measured and is  thought to be significantly lower than  the Galactic disc or halo at the same metallicities \citep[e.g.][]{VanderSwaelmen2013}. The HVS candidate stars mostly display super-solar ratios of [Na/Fe], with one star (Gaia~DR2 1364548016594914560) being closer to solar values. Much like the case of the  $\alpha$ elements, the Na abundances of these HVS candidate stars mostly follow those of the Galactic halo, albeit with a large scatter.
\\
\noindent {\it Al}: \\
Similarly to Na, aluminum (Al) is produced in carbon burning in post main-sequence evolution. However, unlike Na, it is also produced in the MgAl cycle \citep[e.g.][]{Samland1998}. It is mostly dispersed into the interstellar medium via SNII but also through the evolution of AGB stars. The observed ratio of Al to Fe follows the patterns seen for  Mg, Si, Ca, and O, in that at high metallicity (\feh $>$ --1) it increases with decreasing metallicity. This is why Al is often thought of as a `mild' $\alpha$ element. However, unlike the $\alpha$ elements, at lower metallicity [Al/Fe] is observed to decrease with decreasing metallicity, reaching a plateau at sub-solar values, a trend also expected by models \citep[e.g.][]{Kobayashi2006, Nomoto2013}.  We could measure Al in only two of the five observed stars (Fig.~\ref{fig:chem1}) and found, in both cases, that the [Al/Fe] value lies within the scatter of typical halo stars in the metallicity range of --1 $<$ \feh\ $<$ --2~dex. 
\\

\noindent {\it V}: \\
The nucleosynthesis pathway for the production of vanadium (V) is not well understood. This can be seen in the rather poor agreement between the theoretical expectation for the behaviour of [V/Fe] as a function of metallicity and the data \cite[see, for example, Fig.~10 of][]{Nomoto2013}. Regardless of this fact, we can empirically compare the derived values of [V/Fe] for our stars with those from known populations.  Fig.~\ref{fig:chem1} shows the derived values of  [V/Fe] as a function of metallicity for the extreme-velocity stars (black) compared to the Galactic disc \citep[light green circles][]{Battistini2016}, the Galactic halo \citep{Yong2013,Roederer2014}, and the LMC \citep{VanderSwaelmen2013}. The  candidate extreme-velocity stars have a wide range in [V/Fe], again  consistent with the large scatter observed for the Galactic halo.  \\

\noindent {\it Cu}:\\
Copper (Cu) is an odd-Z element that is thought to be largely produced via SNIa and hypernovae  but is also produced by secondary phenomena in massive stars and the weak s-process \citep[e.g.][]{Mishenina2002}. Like many elements, the  exact nucleosynthetic process that forms Cu is still under debate \citep[e.g.][]{Andrievsky2018}. As shown in Fig.~\ref{fig:chem2}, previous surveys found a large scatter in [Cu/Fe] at the metallicities of our stars, and similar low values in the LMC and stellar halo. Again, the stars in our sample fit within the scatter of the stellar halo, but Cu does not allow us to distinguish between stars born in the Galactic disc, halo or LMC. 
\\
\noindent {\it Sc}:\\
Scandium (Sc) is an odd-Z element that, similarly to vanadium, has a poorly understood nucleosynthesis production mechanisms. This is evident by the large discrepancies between the theoretically computed and observed trends of [Sc/Fe] as a function of \feh\ \citep[e.g.][]{Kobayashi2006, Nomoto2013}. As seen in Fig.~\ref{fig:chem2} stars belonging to either the LMC or Galactic disc show [Sc/Fe] increasing with decreasing metallicity, with [Sc/Fe] lower in the LMC at fixed metallicity compared to the Galactic disc \citep{VanderSwaelmen2013}. 

On the other hand, stars in the Galactic halo have [Sc/Fe] values which are either slightly enhanced \citep{Yong2013} or solar \citep{Roederer2014} though with large intrinsic scatter. The candidate extreme-velocity stars are all enhanced in [Sc/Fe] at levels consistent with the Galactic halo but not with the LMC. 

\subsection{Fe-peak elements: Cr, Mn, Co, Ni, Zn}  \label{subsec:Fepeak}
\begin{figure*}
	 \includegraphics[width=2\columnwidth]{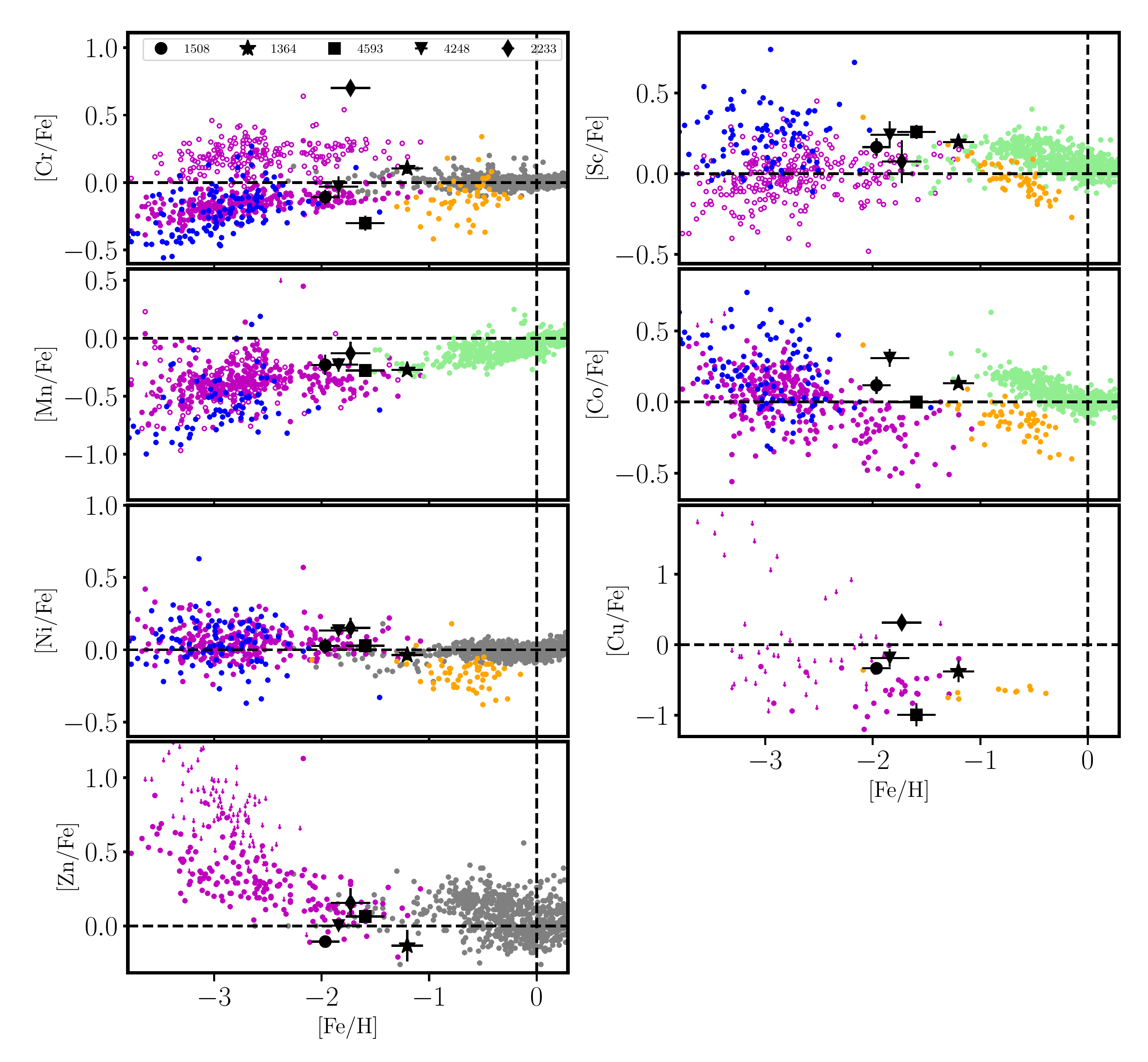}
	\caption{ The [X/Fe] ratios of the program stars (black filled symbols) as a function of \feh\ for Cr, Mn, Ni, and Zn from top to bottom, respectively on the left and Sc, Co, and Cu from top to bottom, respectively on the right. These largely represent the Fe-peak elements. The symbols are the same as in Fig.~\ref{fig:chem1}. }  
	\label{fig:chem2}
\end{figure*}

Unlike the $\alpha$ or odd-Z elements, the Fe-peak elements, such as chromium (Cr), scandium (Sc), manganese (Mn),  cobalt (Co), nickle (Ni), and zinc (Zn), are formed through a variety of paths \citep{Iwamoto1999, Kobayashi2006} but are largely dispersed into the interstellar medium in a similar way as iron. As such, these Fe-peak elements largely track the abundance of Fe. At the metallicity range of the candidate extreme-velocity stars (--2 $<$ \feh\ $<$ --1), [Co/Fe]  and [Zn/Fe] increase with decreasing metallicity while  [Ni/Fe], and [Cu/Fe] remain approximately constant. The other Fe-peak elements (Mn and Cr)  display [X/Fe] values which decrease with decreasing metallicity.

More specifically, Fig.~\ref{fig:chem2} shows that [Cr/Fe] and [Ni/Fe] are approximately constant, at close to the  solar value, for stars in the Galactic disc(s). However, in the Milky Way halo, [Cr/Fe] decreases with decreasing metallicity, with large scatter \citep[e.g.][]{Yong2013, Roederer2014}. It should however be noted  that abundances based on \ion{Cr}{II} measured by \cite{Roederer2014} are offset from the \ion{Cr}{I} abundances of the same authors, which agree with the Cr reported in \cite{Yong2013}. In the disc of the LMC, [Cr/Fe] varies little with metallicity (similar to the lack of trend seen in the Milky Way disc) but at  lower values than in the Galactic disc and with large scatter. The Ni and Cr abundances of the  candidate extreme-velocity stars are consistent with the Galactic halo or, in the case of the most metal-rich of the observed HVS candidates (\gaia\ DR2~1364548016594914560), with the Galactic thick disc or even with the LMC.

Mn is a Fe-peak element and its production is thought to be similar Ni \citep[e.g.][]{Kobayashi2011}, however the observed  chemical trend in the Galactic disc and halo as a function of \feh\ is different than Ni. This unexpected trend is not well understood. However, \cite{Battistini2015}, found that the [Mn/Fe] ratio is constant at low metallicity (rather than decreasing) if NLTE calculations are considered. Since we compute all elemental abundances assuming LTE, we only compare the LTE calculations for Mn (and all elements) from other surveys of the Galactic disc(s), halo and LMC, to our HVS candidate stars. Mn shows a decreasing trend with decreasing metallicity through the Galactic disc(s) and halo. The HVS candidate stars observed in this study follow this expected trend and are consistent with originating in the Milky Way. The HVS candidate stars show trends in each of the Fe-peak elements that are consistent with the Galactic halo (low metallicity and large abundance scatter) in all Fe-peak elements. However, they are inconsistent with the Galactic disc(s) in Fe, Mn and Co and inconsistent with the LMC disc in Fe, Co and Ni.

\subsection{Neutron capture elements: Sr, Y, Zr, Ba, La, Nd, Eu }  \label{subsec:ncap}
\begin{figure*}
 	 \includegraphics[width=2\columnwidth]{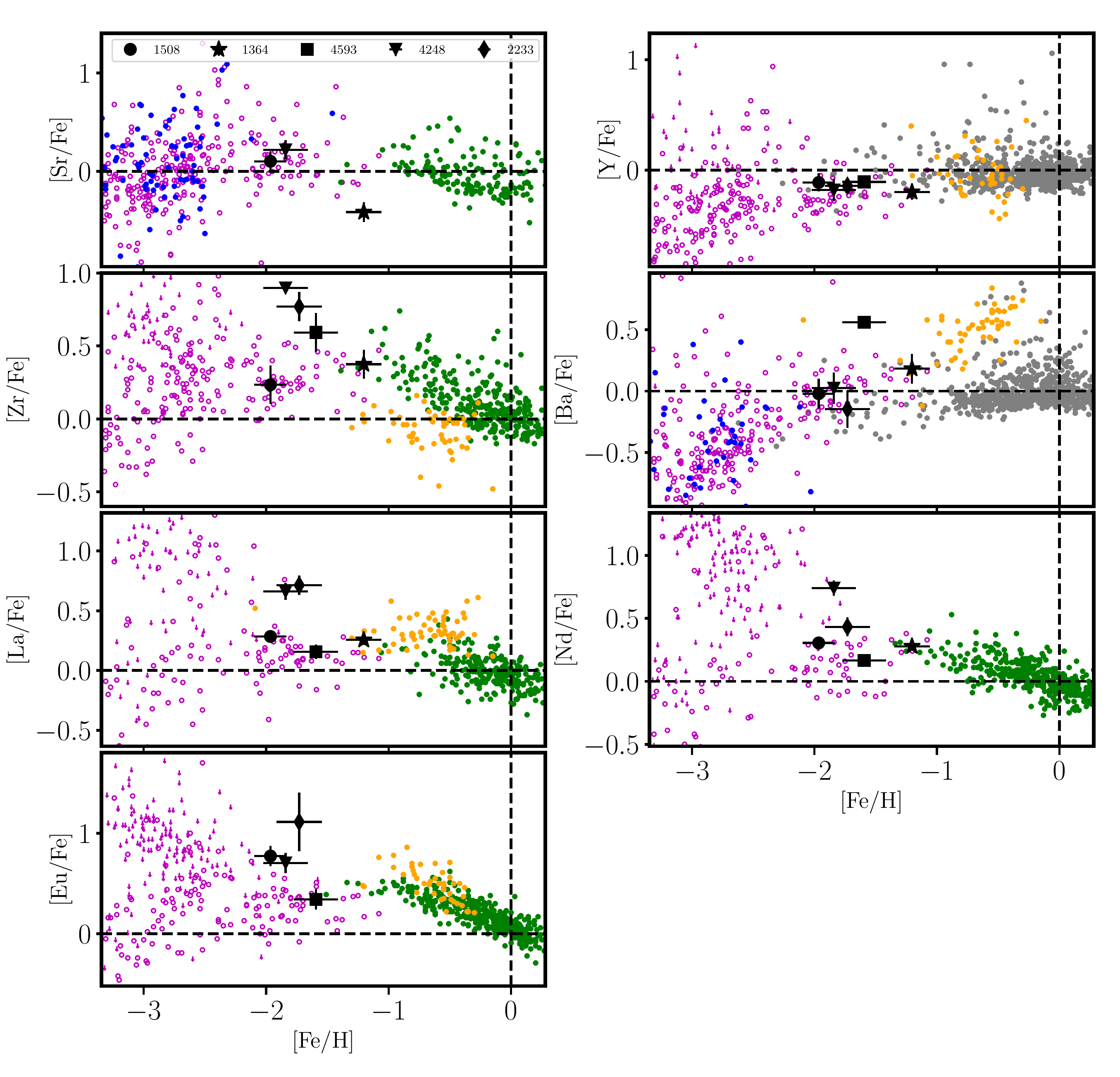}
	\caption{[X/Fe] as a function of metallicity for Sr, Zr, La, and Eu from top to bottom, respectively on the left and Y, Ba, and Nd from top to bottom, respectively on the right.  These largely represent the neutron capture elements. The symbols are the same as Fig.~\ref{fig:chem1} with additional data for the disc taken from \citet[green]{Battistini2015}. }  
	
	\label{fig:chem3}
	\end{figure*}
	
The neutron capture elements are generally split into two separate groups, namely the slow (s-process) and rapid (r-process) neutron capture elements. The s-process elements are often subdivided into a light s-process group comprised of strontium (Sr), yttrium (Y), and zirconium (Zr) and heavy s-process group made of barium (Ba), lanthanum (La), neodymium  (Nd) among others. The exact nucleosynthetic production channel is marginally constrained for some elements such as Ba but not well understood for others (e.g. the r-process element europium, Eu). For example, Ba is largely s-process in origin and is thought to be created primarily in AGB stars and ejected through stellar winds. Other s-process elements (e.g. Sr and Zr) are also thought to be largely created by AGB stars while pure r-process elements (e.g. Eu) may be created by more violent events such as the merger of binary neutron stars. The relative abundances of the neutron-capture elements contain unique insight into the environments of the extreme-velocity stars at their births.

Fig.~\ref{fig:chem3} shows that all the candidate extreme-velocity stars have abundances in the light s-process elements that are consistent with the metal-poor halo (or thick disc for the more metal-rich candidates), showing near solar values in [Sr/Fe] and [Y/Fe] but enhanced in [Zr/Fe]. These values  are, however, inconsistent with those seen in thin disc stars or the disc of the LMC, which would, for example, require both a high metallicity and significantly lower [Zr/Fe]. The candidate extreme-velocity stars are also all consistent with the metal-poor halo (or thick disc) in the heavy s-process elements, showing enhanced values for [La/Fe] and [Nd/Fe] but mostly near solar values for [Ba/Fe]. However, \gaia~DR2 4593398670455374592 has a higher [Ba/Fe] value  and if it were at higher metallicity it could be consistent with stars in the disc of the LMC. The only r-process element we currently measure is Eu and again the extreme-velocity stars lie within the scatter of the stellar halo.

\subsection{Origins of the Hypervelocity Candidate Stars: A Chemical Perspective}
\label{subsec:disucssion}
The elemental abundances we reported above for five of the 28 HVS candidates from \cite{Marchetti2018} which those authors find to have high probability of being unbound to the Milky Way, indicate that these stars do not originate in either the Galactic thin disc or the LMC. This is likely also the case for the HVS candidates that are metal-poor giants found in common with several large spectroscopic surveys. This conclusion is based largely on  the abundances of the $\alpha$  (Mg, Ca, Ti), odd-Z  (Na, V, Sc, ), Fe-peak (Fe, Co, Mn), and neutron capture (Zr, Ba) elements.  Rather, in all elements studied, these fast-moving metal-poor giant stars show abundances typical of Galactic halo stars, with the caveat that the most metal-rich of these stars could have originated in the metal-weak thin disc \citep[e.g.][]{Beers2002,Ruchti2011}, although its high velocity suggests that it may simply be the high-velocity tail of the stellar halo. With the data presented above in Figs.~\ref{fig:chem1}, \ref{fig:chem2}, and \ref{fig:chem3} in mind, we now  put this result into the broader context of the origins of HVS.
\\

\noindent {\it Hills' Mechanism:}\\
One of the primary mechanisms proposed for the production of stars moving well in excess of the Galactic escape velocity is a triple-body encounter between a binary stellar system and a massive object such as  the Milky Way's supermassive black hole or intermediate mass black holes \citep{Hills1988, Yu2003, Gualandris2007}. This mechanism is the expected process  involved in the creation of most of the known and confirmed early-type massive, and hence young, HVSs \citep{Brown2015}. Some of the extreme-velocity stars observed in this work - all late-type, low mass stars - have inferred orbits which are consistent with  a high likelihood of passing within 1~kpc of the Galactic center \citep{Marchetti2018}. If these stars actually did originate close to the Galactic centre and had  been ejected by the purely dynamical Hills' mechanism then their chemical abundance pattern should match that of low-mass stars in the inner Galaxy. Figs~\ref{fig:chem1}, \ref{fig:chem2}, and \ref{fig:chem3} all demonstrate that this is not the case. Low-mass giant stars in the Galactic centre region are  generally  metal-rich (\feh\ $>$ --1.0) and enhanced in the $\alpha$ elements \citep[e.g.][and references therein]{McWilliam2016}. Further, the chemical patterns of these three candidates which have derived orbits that intersect the Galactic centre also do not resemble those of stars in the inner regions of the thin disc. Therefore,  we conclude that it is  very unlikely that these stars have actually originated close to the Galactic centre, a requirement for the Hills' mechanism\footnote{Giant stars have sufficiently large radii that they are  unlikely to be in a tight enough orbit for Hills' mechanism \citep[e.g.][]{Bromley2006, Kenyon2008}.}. \\

\noindent {\it Hypervelocity Stars from the Large Magellenic Cloud or Tidal Debris?:}\\
Another potential origin for HVSs that has been proposed in the literature is as runaway stars from the LMC or tidal debris. This idea was first proposed by \cite{Edelmann2005} who noted that one HVS was less than 20~deg from the centre of the LMC. The idea was then extended by \cite{Boubert2016} and \cite{Boubert2017} who predicted the  kinematic and spatial properties that would result if B-type HVS were runaway stars from the LMC, and found agreement with the observed population. Most recently, \cite{Erkal2018} have shown, using \gaia\ DR2, that at least one of the known and confirmed HVSs does indeed come from the LMC. With this in mind, we assess whether the metal-poor HVS candidates in this work are consistent with an origin in the LMC. 

If these HVS candidate stars were coming from the LMC then (1) their orbits should trace back to the LMC {\it and} (2) their chemistry should resemble the (disc component) of the LMC. \cite{Marchetti2018} has noted that some of their candidates, specifically those with low probability of crossing the Galactic disc (P$_{MW} <$ 0.50), may have an extragalactic origin, such as being ejected from the LMC or tidally stripped debris. Again the detailed chemical abundance distributions provide a means to distinguish between an origin in the Milky Way and an extragalactic origin.  We determine whether the chemical distribution of the HVSs candidate stars are consistent with the LMC in Figs~\ref{fig:chem1}, \ref{fig:chem2}, and \ref{fig:chem3}. It is clear  that the HVSs have chemical abundance signatures that do not match those of the LMC in various elements, including Na, Mg, Ca Sc, Co, Ni, and Zr. More specifically, the candidate extreme-velocity stars  are more enhanced in these elements than are stars in  the LMC. This rules out an origin in the LMC.  Tidal debris from other satellite systems could have low metallicity, as observed for the candidate HVSs, but would be expected to have the low values of the abundance ratios [Mg, Si, Ca, Ti/Fe] found in old, low-mass stars in present-day dwarf galaxies, in contrast to the enhanced values we observe. 

\noindent {\it Are these Stars Really Unbound?:}\\
\cite{Hattori2018}  compute the production rates of HVSs from globular clusters, through either ejection of the binary companion in a Type Ia supernova or the interaction between a binary star system and an intermediate mass black hole. Interestingly, they found that the rate predicted  is a couple of orders of magnitude too low to replenish the population of HVSs should they actually escape from the Galaxy. This shortfall led them to conclude that, regardless of how the HVS candidates achieved their high velocities, they are in fact not escaping, but are bound to the Galaxy. Our results add a new dimension, that of the chemical signature, to this argument.

Firstly, their chemical composition indicates that the extreme velocity stars studied here most closely resembles the Galactic halo. In addition, some stars, generally referred to as second generation, in globular clusters are known to display an anti-correlation between Mg and Al \citep[e.g.][]{Bastian2018}. This would make these globular cluster stars distinguishable from field stars because they have high [Al/Fe] but low [Mg/Fe], thereby having a significantly higher [Al/Mg] ratio compared to field stars. However, for two of the three stars where Al is measured, the [Al/Fe] is not high and the [Mg/Fe] is enhanced, suggesting that these stars may just be regular field halo or are former members of the first-generation of a globular cluster, which do not show a Mg-Al anti-correlation.

Furthermore, in both HVSs candidate samples from \cite{Hattori2018} and \cite{Marchetti2018}, there are similar numbers of extreme-velocity stars with positive and negative RVs. All of  these points taken together indicate that these HVSs candidate stars are simply just typical bound halo stars which are at the high speed tail of the velocity distribution of the stellar halo. \

\section{Summary} 
\label{sec:summary}
Hypervelocity stars are intriguing objects that are moving faster than the escape speed of the Milky Way at their observed location and are thus unbound. These rare stars were first proposed as the `smoking-gun' evidence for a super-massive black hole at the Galactic Centre by  \cite{Hills1988} and  massive, young HVSs were first discovered observationally in 2005 \citep{Brown2005}. With exquisite astrometric data from the \gaia\ second data release, more than 40 late-type (FGK-type) hypervelocity star candidates have been proposed \citep{Marchetti2018, Boubert2018, Hattori2018}. In this paper, we report the first spectroscopic followup and chemical characterisation of a subsample of newly proposed HVS candidates. We observed five HVS candidates from \cite{Marchetti2018} with the high-resolution ARCES spectrograph on the 3.5m telescope at the Apache Point Observatory. The observational properties of these stars can be found in Table~\ref{tab:obsprops}. In addition to these observed targets, four more were found to be serendipitously observed by the RAVE, APOGEE, and LAMOST spectroscopic surveys.

The spectra from the APO were processed in the standard way (see section~\ref{sec:method} for more details). Our first aim was to confirm the RV of these fast-moving stars, which are often very large (RV $\geq$ 300~\kms). On average the RV measured in our work using APO and that from \gaia\ agree to within $\sim$ 1~\kms\ with a dispersion of 2~\kms. This encouraging result indicates that \gaia\ is able to measure accurate RVs, including high RV, for stars as dim as G~$\sim$~13.5~mag.

We then completed an extensive stellar atmospheric and chemical characterisation of the five observed HVS candidates and found that all five HVS candidates are metal-poor giants. This implies a dramatically different parent population from that of the known massive O- and B-type HVSs. Further, the HVS candidate stars that we identified as serendipitously observed in published large surveys are also metal-poor giant stars though their chemical detailed chemistry is unknown in most cases. The chemical abundance patterns we derived for the five stars observed with APO comprised 22 species, including Mg, Ti, Si, Ca ($\alpha$ group), Fe, Ni, Co, Cr, Sc, Mn (Fe-peak), Na, Al, V, Cu (odd-Z group), and Sr, Y, Zr, Ba, La, Nd, Eu (Neutron capture). In all chemical elements studied, the HVS candidate stars match typical Galactic halo stars and do not resemble stars in any of the inner Galactic bulge/disc, outer disc near the Sun or the LMC. We conclude that these stars are most likely just bound halo stars that are at the tail of its velocity distribution.  A similar conclusion was reached by \cite{Boubert2018} on pure dynamical grounds. These results suggest that the Milky Way escape velocity is higher than previously estimated \citep{Piffl2014, Williams2017}, and  consistent with the recent studies incorporating data from \gaia\ DR2 \citep[e.g.][Monari et al. in preparation]{Hattori2018}. In turn this implies a higher total mass for the Milky Way in agreement with recent analyses of the orbits of globular clusters using \gaia\ DR2 astrometric data \citep[e.g.][]{Watkins2018, Posti2018}. We note that these higher total (dark matter) masses imply increased dark matter substructure and higher numbers of satellite galaxies \citep{Fattahi2016}, with obvious ramifications for the `missing satellite' issue \citep[e.g.][]{Bullock2017} 

The identification of HVS is both timely, given the release of the exquisite \gaia\ astrometric data, and necessary, given our understanding of the supermassive black hole and surrounding stellar population in  the Galactic centre region. The discovery of many stars with extreme kinematics  will ultimately allow us to understand the parent population of HVS to constrain the production mechanism(s) that dominates their creation.
The results presented here demonstrate the power of combining {\it complementary} approaches of dynamical orbit integration analysis and chemical characterisation to constrain the origins of HVSs and ultimately to provide a better understanding of these rare stars and the physical mechanisms that produce them.

\appendix{}
\section{Online Tables}
We provide an online table, a section of which is shown in Table~\ref{tab:online} for reference. This table includes the line-by-line abundances for each star, line and elemental species. The table also includes the atomic data, specifically the oscillator strength ($\log{gf}$), the wavelength (\AA), and excitation potential (in eV) for each line.

\begin{table*}
\caption{Line-By-Line Chemical Abundances and Atomic Information}
\label{tab:online}
\centering 
\begin{tabular}{l l l l l l l l l l l  }
\hline\hline
Star&Element&$\lambda$&$\log(gf)$&Excitation Pot.& $\log(\epsilon)$\\
 & & (\AA) &  & (eV) &   \\

\hline 
\hline

1508756353921427328&Ca I&5260.38&-1.719&2.521&4.949\\
1508756353921427328&Ca I&5261.70&-0.579&2.521&4.678\\
1508756353921427328&Ca I&5349.46&-0.310&2.709&4.863\\
1508756353921427328&Ca I&5512.98&-0.464&2.933&4.900\\
1508756353921427328&Ca I&5581.96&-0.555&2.523&4.766\\
1508756353921427328&Ca I&5588.74&0.358&2.526&4.311\\
1508756353921427328&Na I&5682.63&-0.706&2.102&4.091\\
1508756353921427328&Na I&5688.20&-0.404&2.104&4.462\\
1508756353921427328&Na I&6160.74&-1.246&2.104&4.375\\
1508756353921427328&Na I&8183.25&0.237&2.102&4.547\\
1364548016594914560&Na I&4497.65&-1.574&2.104&5.068\\
1364548016594914560&Na I&5682.63&-0.706&2.102&4.827\\
1364548016594914560&Na I&5688.20&-0.404&2.104&4.881\\
1364548016594914560&Na I&6160.74&-1.246&2.104&4.784\\
...&...&...&...&...&...\\

\hline  \hline     
\end{tabular}
\\
\raggedright
The \gaia\ DR2 source id is tabulated in column 1. The element, wavelength (\AA), oscillator strength ($\log(gf)$), and excitation potential (in eV) for each absorption feature consider in each star is in column 2, 3, 4, and five, respectively. The last column of the table indicates the derived abundance for each absorption feature in each star.
\end{table*}

\section*{Acknowledgements}
{\small 
We thank the referee for their detailed and constructive comments that improved this manuscript. K.H. would like to thank Kohei Hattori, Nathan Leigh, Marwan Gerbran, Douglas Boubert, James Guillochon, and Idan Ginsburg for lively discussions. RFGW thanks her sister, Katherine Barber, for her support. RFGW also thanks the Leverhulme Trust for a Visiting Professorship at the University of Edinburgh, held while this work was being completed. We would also like to thank Scott Kenyon for his detailed comments. Based on observations obtained with the Apache Point Observatory 3.5-meter telescope, which is owned and operated by the Astrophysical Research Consortium. We thank all at APO for their assistance during remote observing. This project was developed in part at the 2018 NYC Gaia Sprint, hosted by the Center for Computational Astrophysics of the Flatiron Institute in New York City.

This work has made use of data from the European Space Agency (ESA)
mission {\it Gaia} (\url{https://www.cosmos.esa.int/gaia}), processed by
the {\it Gaia} Data Processing and Analysis Consortium (DPAC,
\url{https://www.cosmos.esa.int/web/gaia/dpac/consortium}). Funding
for the DPAC has been provided by national institutions, in particular
the institutions participating in the {\it Gaia} Multilateral Agreement.

\bibliography{bibliography}
\bsp	
\label{lastpage}
\end{document}